\documentclass[preprint,showpacs,showkeys,preprintnumbers,amsmath,amssymb]{revtex4}
 \usepackage{dcolumn}% Align table columns on decimal point
 \usepackage{bm}% bold math

\begin{document}

 \title{Vector and Spinor Decomposition of SU(2) Gauge Potential, their Equivalence  and
 Knot Structure in SU(2) Chern-Simons Theory}

 \author{DUAN Yi-shi}
 \author{REN Ji-rong}
 \author{LI Ran}
 \thanks{Correspondence author. Email: liran05@st.lzu.edu.cn}
 \affiliation{Institute of Theoretical Physics,  Lanzhou University, Lanzhou, 730000, China}

 \date{\today}

 \begin{abstract}
 In this paper, spinor and vector decomposition of SU(2) gauge potential
 are presented and their equivalence is constructed using a simply proposal.
 We also obtain the action of Faddeev nonlinear $O(3)$ sigma model from the
 SU(2) massive gauge field theory which is proposed according to the gauge
 invariant principle. At last, the knot structure in SU(2)
 Chern-Simons filed theory is discussed in terms of the
 $\phi$--mapping topological current theory.
 The topological charge of the knot is characterized
 by the Hopf indices and the Brouwer
 degrees of $\phi$-mapping.
 \end{abstract}



 \maketitle

 \section{Introduction}

 Since the decomposition theory of gauge potential reveals the inner
 structure of gauge potential, it inputs the geometrical and
 topological information to the gauge potential (i.e. the connection of principal
 bundle), and establishes a direct relationship between differential
 geometry and topology of gauge field.
 The above viewpoint of inner-structure is expected to enrich the
 gauge theory with deeper physical contents; actually an outstanding
 case in point is the general relativity. In general relativity or
 4-dimensional Riemannian geometry, the connection can
 be expressed in terms of the fundamental field or metric
 $g_{\mu\nu}$ which is necessary to be introduced to describe the
 gravity field.
 In recent years the decomposition
 theory of gauge potential has
 played a more and more important role in theoretical physics and
 mathematics.  From this viewpoint much
 progress has been made by other authors\cite{faddeev,cho} and by us, such as the
 decomposition of U(1) gauge potential and U(1) Chern-Simons,
 the decomposition of SU(2) connection
 and the Skyrme theory,
 the decomposition of SU(N) connection
 and the effective theory of SU(N) QCD,
 the decomposition of SO(N) spin connection and
 the structure of GBC topological current
 \cite{faddeev,cho,duanprd,sun,pbranes,spinordecom,choskyme}.

 Cho\cite{choskyme} derive a generalized Skyrme action
 from the Yang-Mills action of SU(2) QCD,
 which is proposed to be an effective action of SU(2) QCD in the infrared
 limit.
 This stimulate us to explore the relationship between SU(2)
 Yang-Mills theory and the Faddeev nonlinear $O(3)$ sigma model
 which has intriguing consequences\cite{sigma}.
 One of the co-authors Prof.Duan\cite{gaugedecom} have pointed out almost twenty years
 ago that
 gauge potential $A_\mu$ can be decomposed into two parts: $a_\mu$
 and $b_\mu$.
 Here $a_\mu$ satisfies the gauge transformation $a_\mu'=ga_\mu g^{-1}+\partial_\mu gg^{-1}$,
 and the $b_\mu$ satisfies the
 adjoint transformation $b_\mu'=gb_\mu g$.
 The $a_\mu$ part may show the
 geometry property of system and the $b_\mu$ part may be looked upon as
 vector boson which would be massive without introducing Higgs mechanism
 and spontaneous symmetry breaking. From this point of view,
 we can introduce a massive term in terms of $b_\mu$ part
 to construct a massive gauge field theory without destroying the
 gauge invariance which can naturally deduce the
 action of Faddeev nonlinear $O(3)$ sigma model using the vector
 decomposition of SU(2) gauge potential.

 In this paper, spinor and vector decomposition of SU(2) gauge potential
 are presented and their equivalence is constructed using a simply proposal.
 We also obtain the action of Faddeev nonlinear $O(3)$ sigma model from the
 SU(2) massive gauge field theory which is proposed according to the gauge
 invariant principle. At last, the knot structure in SU(2)
 Chern-Simons filed theory is discussed in terms of the
 $\phi$--mapping topological current theory proposed by Prof.Duan. The topological charge
 of the knot is characterized by the Hopf indices and the Brouwer degrees of $\phi$-mapping.

 \section{decomposition of SU(2) connection}

 We begin with a brief review of our previous work\cite{spinordecom} on
 the spinor
 decomposition of SU(2) connection. Let $M$ be a compact oriented $4$-dimensional manifold,
 on which the principal bundle $P(\pi, M, SU(2))$ is defined. It is well
 known that in the SU(2) gauge field theory with spinor
 representation $\Psi$  , the covariant derivative is defined as
 \begin{equation}
 D_{\mu}\Psi=\partial_{\mu}\Psi-\frac{1}{2i}A^a_\mu\sigma^a\Psi,
 \end{equation}
 where
 \begin{equation}
 A=A_\mu dx^\mu=\frac{1}{2i}A^a_\mu\sigma^a dx^\mu
 \end{equation}
 is the SU(2) gauge potential, i.e. the connection of principle
 bundle $P$, and $T^a=\frac{1}{2i}\sigma^a(a=1,2.3)$ are the SU(2)
 generator with $\sigma^a$ being Pauli matrix.
 The final result of spinor decomposition is
 \begin{eqnarray}
 A^a_\mu&=&\frac{i}{\Psi^{\dagger}\Psi}
 (\Psi^{\dagger}\sigma^a\partial_\mu\Psi
 -\partial_\mu\Psi^{\dagger}\sigma^a\Psi)\nonumber\\
 & &-\frac{i}{\Psi^{\dagger}\Psi}
 (\Psi^{\dagger}\sigma^aD_\mu\Psi
 -D_\mu\Psi^{\dagger}\sigma^a\Psi).
 \end{eqnarray}
 The traditional decomposition theory of gauge potential
 always uses the parallel field
 condition $D_\mu\Psi=0$ and the normalized spinor $\Psi$ with $\Psi^\dagger\Psi=1$,
 so the vectorial transformation part of
 $A_\mu^a$ disappears and $A_\mu^a$ can be expressed as
 \begin{equation}
 A^a_\mu=i
 (\Psi^{\dagger}\sigma^a\partial_\mu\Psi
 -\partial_\mu\Psi^{\dagger}\sigma^a\Psi),
 \end{equation}
 which satisfies the SU(2) gauge transformation.
 The expression (4) is useful to reveal the inner
 structure of the second Chern class.
 Noticing the index $a$ in gauge potential $A^a_\mu$, the gauge potential $A^a_\mu$
 can also be described as the vector form $\vec{A}_\mu$.

 Another method of decomposing the SU(2) connection has been done by
 Cho\cite{cho} and Faddeev\cite{faddeev} in terms of the vector
 topological
 field $\vec{n}$ which is taken as the basic field on the manifold
 $M$. The covariant derivative of $\vec{n}$ is defined as
 \begin{equation}
 D_\mu\vec{n}=\partial_\mu\vec{n}-\vec{A}_\mu\times \vec{n}.
 \end{equation}
 The decomposition formula of $\vec{A}_\mu$ is
 \begin{equation}
 \vec{A}_\mu=C_\mu\vec{n}+\vec{n}\times\partial_\mu\vec{n}+\vec{X}_\mu,
 \end{equation}
 where $C_\mu=\vec{A}_\mu\cdot\vec{n}$ is the projection of gauge
 potential $\vec{A}_\mu$ on $\vec{n}$ direction and $\vec{X}_\mu=\vec{n}\times
 D_\mu\vec{n}$ which is perpendicular with $\vec{n}$.
 Observed that $\vec{n}$ represents a direction which
 is purely associated with orientation of the moving frame
 $\{\vec{n}, \partial_\mu\vec{n}, \vec{n}\times\partial_\mu\vec{n}\}$, one
 can express $\vec{X}_\mu$ in terms of the moving frame as
 \begin{equation}
 \vec{X}_\mu=f_1\partial_\mu\vec{n}+f_2\vec{n}\times\partial_\mu\vec{n}.
 \end{equation}
 The decomposition of SU(2) connection (3) and (6) should be
 equivalent. The topological field $\vec{n}$
 which selects the color direction at each space-time
 point\cite{choskyme} can be constructed using the spinor field like
 this
 \begin{equation}
 n^a=\Psi^{\dagger}\sigma^a\Psi.
 \end{equation}
 Once the relationship between the topological filed $\vec{n}$ and
 the spinor field $\Psi$ is set up, the projection of gauge
 potential $C_\mu$
 can be calculated  where
 \begin{equation}
 C_\mu=A_\mu^an^a=2i\Psi^{\dagger}\partial_\mu\Psi,
 \end{equation}
 and the decomposition formula (4) can be directly driven from (6) under the
 gauge parallel condition $D_\mu\vec{n}=0$ after a simple calculation.
 Then, two kinds of the decomposition of the SU(2) connection is
 proved to be equivalent.

 One of the co-authors Prof.Duan have pointed out almost twenty years ago that
 gauge potential should be decomposed in terms of the gauge
 covariant $A_\mu=a_\mu+b_\mu$,
 which the $a_\mu$ satisfies the gauge transformation $a_\mu'=ga_\mu g^{-1}+\partial_\mu gg^{-1}$,
 and the $b_\mu$ satisfies the
 adjoint transformation $b_\mu'=gb_\mu g$.
 The $a_\mu$ part may show the
 geometry property of system and the $b_\mu$ part may be looked upon as
 vector boson which would be massive.

 Under the infinitesimal gauge transformation\cite{cho}
 \begin{equation}
 \delta\vec{n}=-\vec{\alpha}\times\vec{n},\;\;
 \delta\vec{A}_\mu=D_\mu\vec{\alpha},
 \end{equation}
 one have
 \begin{equation}
 \delta C_\mu=\vec{n}\cdot\partial_\mu\vec{\alpha},\;
 \delta \vec{X}_\mu=-\vec{\alpha}\times\vec{X}_\mu.
 \end{equation}
 $\vec{X}_\mu$ part in decomposition formula (6) transforms
 covariantly under the gauge transformation.
 Then we can introduce SU(2)
 massive gauge field theory which the Lagrange is defined like this without
 destroying the gauge covariant
 \begin{equation}
 {\cal{L}}=-\frac{1}{4}\vec{F}_{\mu\nu}\cdot\vec{F}_{\mu\nu}+\lambda
 \vec{X}_\mu^2.
 \end{equation}
 One can simply calculate the massive term in the above Lagrange from (7)
 which gives
 \begin{equation}
 \lambda\vec{X}_\mu^2=\lambda(f^2_1+f^2_2)\partial_\mu\vec{n}^2.
 \end{equation}
 Noticing that this massive term is just the first term in the Faddeev's
 model which the action is defined as
 \begin{equation}
 S=\int
 d^4x\{m^2(\partial_\mu\vec{n})^2
 +\frac{1}{e^2}(\vec{n}\cdot\partial_\mu\vec{n}\times\partial_\nu\vec{n})^2\},
 \end{equation}
 and the second term named Faddeev-Skyrme term is so familiar that we are
 not to discuss in detail, so one
 can really deduce the action of Faddeev's model which is a unique
 action for describing SU(2) Yang-Mills theory at low energies from
 the Lagrange of SU(2) massive gauge field.
 The massive gauge field theory constructed here
 may be a suitable theory for describing the
 interaction of elementary particle. A further discussion of
 the quantization and renormalization of massive gauge filed theory will be published
 elsewhere.

 \section{knot structure in SU(2) Chern-Simons action}

 Chern-Simons action is a very
 important topological invariant which have deep relationship
 with the knot invariant as pointed out by E.Witten in his pioneer
 work\cite{witten}. In fact, Duan has pointed out in \cite{duanprd}
 that U(1) Chern-Simons action is an important
 invariant required to describe the topology of knot in
 Chern-Simons field theory. Here, we will study the
 knot structure in SU(2) Chern-Simons field theory.

 Chern-Simons action is the integral of the Chern-Simons
 $3-$form\cite{chern}
 \begin{equation}
 S=\frac{1}{8\pi^2}\int_V Tr(A\wedge dA-\frac{2}{3}A\wedge A\wedge
 A),
 \end{equation}
 where $V$ is the space volume.
 It can also be expressed as
 \begin{equation}
 S=-\frac{1}{16\pi^2}\int_V \epsilon^{\mu\nu\lambda}(\vec{A}_\mu\cdot\partial_\nu\vec{A}_\lambda
 -\frac{1}{3}\vec{A}_\mu\cdot\vec{A}_\nu\times\vec{A}_\lambda)d^3x.
 \end{equation}
 Using the decomposition formula (6) and the
 condition $D_\mu\vec{n}=0$, one can easily drive at
 \begin{equation}
 S=-\frac{1}{16\pi^2}\int_V
 (\epsilon^{\mu\nu\lambda}C_\mu\partial_\nu C_\lambda+
 \epsilon^{\mu\nu\lambda}C_\mu H_{\nu\lambda})d^3x,
 \end{equation}
 where
 $H_{\mu\nu}=\vec{n}\cdot\partial_\mu\vec{n}\times\partial_\nu\vec{n}$
 plays an essential role in studying the knot structure in SU(2) Chern-Simons field theory.
 The normalized two component spinor $\Psi$ can be expressed by
  \begin{equation}\label{}
 \Psi=
 \left(
   \begin{array}{c}
     l^0+il^1 \\
     l^2+il^3 \\
   \end{array}
 \right)\;\;,
 \end{equation}
 where $l^a(a=0, 1, 2, 3)$ is a real unit vector.
 After some algebra, the first term in Eq.(17) is
 \begin{equation}
 S^{(1)}=\frac{1}{12\pi^2}\int\epsilon_{abcd}
 \epsilon^{\mu\nu\lambda}l^a\partial_\mu
 l^b\partial_\nu l^c \partial_\lambda l^d d^3x,
 \end{equation}
 which is just the winding number of Gauss mapping $S^3\mapsto S^3$.
 Under SU(2) gauge transformation, the Chern-Simons action
 transforms like this
 \begin{equation}
 S'=S+m,
 \end{equation}
 where
 \begin{equation}
 \omega=\int_V Tr(g^{-1}dg\wedge g^{-1}dg\wedge g^{-1}dg)
 \end{equation}
 is the winding number of map $S^3\mapsto SU(2)$ with $g$
 being an element of SU(2).
 To see more clearer, $g$ can be parameterized as
 \begin{equation}
 g=l^a s^a,\;\;s^a=(I,i\vec{\sigma}).
 \end{equation}
 Then one exactly get Eq(19) from (21).
 So the firse term in Chern-Simons action (17) can be wiped off after
 gauge transformation. Now we focus on the second term
 of Eq.(17)
 \begin{equation}
 S=\frac{1}{16\pi^2}\int_V \epsilon^{\mu\nu\lambda}C_\mu H_{\nu\lambda}d^3x.
 \end{equation}

 We derive the 2-dimensional topological current from the field
 tensor $H_{\mu\nu}$, and show that there are knot structures inhering
 in SU(2) Chern-Simons. Int fact, the unit vector $\vec{n}$ is
 the section of sphere bundle $S^2$. Defining a 2-component vector
 $\vec{\phi}=(\phi^1,\phi^2)$ on this $S^2$, i.e.
 $\tilde\phi\cdot\vec{n}=0(\tilde\phi=\phi^a/\|\phi\|,a=1,2)$,
 it can be proved that\cite{wuyang}
 \begin{equation}
 H_{\mu\nu}=2\epsilon_{ab}\partial_\mu\tilde\phi^a\partial_\nu\tilde\phi^b.
 \end{equation}
 According to $\phi$-mapping topological current theory\cite{duanphimap},
 the 2-dimensional topological current is defined as
 \begin{equation}
 j^{\lambda}=\frac{1}{4\pi}\epsilon^{\lambda\mu\nu}\epsilon_{ab}
 \partial_\mu\tilde\phi^a\partial_\nu\tilde\phi^b,
 \end{equation}
 so we have
 \begin{equation}
 j^{\lambda}=\frac{1}{8\pi}\epsilon^{\lambda\mu\nu}H_{\mu\nu}.
 \end{equation}

 Then using $\partial_\mu\tilde\phi^a=\partial\phi^a/\|\phi\|+\phi^a/\partial_\mu(1/\|\phi\|)$
 and the Green function formula in $\phi-$space $\partial_a\partial_a
 ln\|\phi\|=2\pi\delta^2(\vec{\phi})$ with
 $\partial_a=\partial/\partial\phi^a$,
 it can be proved that
 \begin{equation}
 j^{\lambda}=\delta^2(\phi)D^\lambda(\frac{\phi}{x}),
 \end{equation}
 where
 \begin{equation}
 D^\lambda(\frac{\phi}{x})=\frac{1}{2}
 \epsilon^{\lambda\mu\nu}\epsilon_{ab}\partial_\mu\phi^a\partial_\nu\phi^b
 \end{equation}
 is the Jacobian vector.This expression of $j^\lambda$ provides an
 important conclusion
 \begin{equation}
 j^\lambda
 \begin{cases}
 =0& \text{if and only if $\vec{\phi}\neq 0$}, \\
 \neq 0& \text{if and only if $\vec{\phi}=0$}.
 \end{cases}
 \end{equation}
 So it is necessary to study the zero points of $\vec{\phi}$ to
 determine the nonzero solution of $j^\lambda$.  The implicit function
 theory\cite{implicit} show that under the regular condition
 \begin{equation}
 D^\lambda(\frac{\phi}{x})\neq 0,
 \end{equation}
 the general solutions of
 \begin{equation}
 \phi^1(x^1, x^2, x^3)=0, \\
 \phi^2(x^1, x^2, x^3)=0,
 \end{equation}
 can be expressed as
 \begin{equation}
 x^1=x^1_k(s), x^2=x^2_k(s), x^3=x^3_k(s).
 \end{equation}
 which represent $N$ isolated singular strings $L_k(k=1, 2, \ldots, N)$
 with string parameter $s$.
 In terms of the viewpoint of topological defect, the vector function $\vec{\phi}$ is
 just the orderparameter of the defects and these singular strings are just the topological
 defects.

 In $\delta$-function theory\cite{delta},  one can prove that in three dimension
 space
 \begin{equation}
 \delta^2(\vec{\phi})=\sum_{k=1}^{N}\beta_{k}\int_{L_k}\frac{\delta^3(\vec{x}-\vec{x}_k(s))}{\mid
 D(\frac{\phi}{u})\mid_{\Sigma_k}}ds,
 \end{equation}
 where
 $D(\frac{\phi}{u})=\frac{1}{2}\epsilon^{\mu\nu}\epsilon_{ab}\frac{\partial\phi^a}{\partial
 u^\mu}\frac{\partial\phi^b}{\partial u^\nu}$ and $\Sigma_k$ is the $k$th
 planar element transverse to $L_k$ with local coordinates
 $(u^1, u^2)$. The positive integer $\beta_k$ is the Hopf index of
 $\phi$-mapping, which means that when $\vec{x}$ covers the neighborhood of the
 zero point $\vec{x}_k(s)$ once, the vector field $\vec{\phi}$ covers the
 corresponding region in $\phi$ space $\beta_k$ times.  Meanwhile
 from Eq.(33), one have
 \begin{equation}
 \partial_\mu\phi^adx^\mu|_{L_k}=0,
 \end{equation}
 then the tangent vector of $L_k$ is given by
 \begin{equation}
 \frac{dx^\lambda}{ds}\mid_{L_k}=\frac{D^\lambda(\frac{\phi}{x})}{D(\frac{\phi}{u})}\mid_{L_k}.
 \end{equation}
 Then the inner topological structure of $j^\lambda$ is
 \begin{equation}
 j^\lambda=\sum_{k=1}^N
 W_k\int_{L_k}\frac{dx^\lambda}{ds}\delta^3(\vec{x}-\vec{z_k}(s))ds,
 \end{equation}
 where $W_k=\beta_k\eta_k$ is the winding number of $\vec{\phi}$
 around $L_k$,  with $\eta_k=sgnD(\frac{\phi}{u})\mid_{\Sigma_k}=\pm 1$
 being the Brouwer degree of $\phi$ mapping. The topological charge
 of the defect line $L_k$ is
 \begin{equation}
 Q_k=\int_{\Sigma_k}j^\lambda d\sigma_\lambda=W_k.
 \end{equation}
 Using Eq.(25) and (38), the part of SU(2) Chern-Simons action that
 we are care for is expressed as
 \begin{equation}
 S=\frac{1}{2\pi}\int_V C_\lambda j^\lambda d^3x=\frac{1}{2\pi}\sum_{k=1}^l W_k\int_{L_k}
 C_\lambda dx^\lambda.
 \end{equation}
 It can be seen that when these singular strings are closed curves or
 more generally are a family of $N$ knots $\gamma_k(k=1, 2, \ldots, N)$,
 the inner structure of topological current is
 \begin{equation}
 j^\lambda=\sum_{k=1}^N
 W_k\oint_{\gamma_k}\frac{dx^\lambda}{ds}\delta^3(\vec{x}-\vec{z_i}(s))ds,
 \end{equation}
 and SU(2) Chern-Simons action is
 \begin{equation}
 S=\frac{1}{2\pi}\sum_{k=1}^l W_k\oint_{\gamma_k}
 C_\lambda dx^\lambda.
 \end{equation}
 Consider the infinitesimal gauge transformation
 $\vec{\alpha}=\alpha\vec{n}$, then $C_\lambda$ transforms like
 this
 \begin{equation}
 C_\lambda'=C_\lambda+\partial_\lambda\alpha,
 \end{equation}
 which is just the U(1) gauge transformation.
 It is seen that the $\partial_\lambda\alpha$ term in Eq.(43) contributes
 nothing to the integral in Eq.(42). Hence the expression (42) is invariant
 under the infinitesimal gauge transformation along the direction of $\vec{n}$.
 These closed singular strings are just the knot structure in SU(2)
 Chern-Simons field theory.

 At last, we must point out that, in this section, we
 have used the regular condition $D^\lambda(\phi /x)\neq
 0$. Generally, this condition is not always tenable. When
 this condition fails, branch process will occur. A further
 study of the branch process will appeared in our future paper.

 \section*{ACKNOWLEDGEMENT}

 This work was supported by the National Natural Science Foundation
 of China and the Doctoral Foundation from the Ministry of Education
 of China.

 \end{document}